\documentclass[]{raa}            
\usepackage{graphicx,times}
\usepackage{natbib}
\usepackage{epstopdf}        

\providecommand{\bm}[1]{\mbox{\boldmath$#1$\unboldmath}}
\def\kms{$\rm{km~s}^{-1}$}
\def\etal{et al.~}

\def\Oiii{[O{\sc~iii}]}
\def\Oii{[O{\sc~ii}]}

\def\Hb{H{\sc$\beta$}}

\begin{document}

   \title{The quasar SDSS J142507.32+323137.4 : dual AGNs?}

 \volnopage{ {\bf 20xx} Vol.\ {\bf x} No. {\bf XX}, 000--000}
   \setcounter{page}{1}
 
   \author{Zhixin Peng      \inst{1,2}
   \and Yanmei Chen       \inst{1,2}
   \and Qiusheng Gu       \inst{1,2}
   \and Chen Hu     \inst{3}
   }
 
   \institute{ Department of Astronomy, Nanjing University, Nanjing 210093, China; {\it zxpeng@nju.edu.cn}
\and
Key Laboratory of Modern Astronomy and Astrophysics (Nanjing University), Ministry of Education, Nanjing 210093, China \\
\and
Key Laboratory for Particle Astrophysics, Institute of High Energy Physics, Chinese Academy of Sciences,
19B Yuquan Road, Beijing 100049, China\\
\vs \no
   {\small  Received [year] [month] [day]; accepted [year] [month] [day] }
}
\abstract{ We analyze the optical spectrum of type 1 QSO SDSS J1425+3231. This object is interesting 
since its narrow emission lines such as \Oiii$\lambda\lambda$4959, 5007 are double-peaked, and the 
line structure can be modeled well by three Gaussian components: two components for the two peaks 
(we refer the peaks at low/high redshift as ``the blue/red component") and 
another one for the line wing which has the same line center as that of the blue component, but $\sim 3$ 
times broader. The separation between the blue and red components is $\sim 500$ km/s with blue
component $\sim 2$ times broader than the red one. The \Hb\ emission can be separated into 
four components: two for the double-peaked narrow line and two for the broad 
line which comes from the broad line region (BLRs). The black hole mass estimated from the broad 
\Hb\ emission line using the typical reverberation mapping relation is  $0.85\times10^{8}M_\odot$,
which is consistent with that derived from parameters of  \Oiii$\lambda$ 5007 of the blue component.
We suggest this QSO might be a dual AGN system, the broad \Hb\ emission line is mainly 
contributed by the primary black hole (traced by the blue component) while the broad \Hb\ component 
of the secondary black hole (traced by the red component) is hard to be separated out considering a
resolution of $\sim$2000 of SDSS spectra or it is totally obscured by the dusty torus.
\keywords{galaxies: active --- galaxies: individual (SDSS J142507.32+323137.4) --- quasars: emission lines }
}

   \authorrunning{Z.X. Peng, Y.M. Chen \& Q.S. Gu }             
   \titlerunning{SDSS J142507.32+323137.4
 }  
   \maketitle

%
%
\section{Introduction}           
\label{sect:intro}
Early in 1980s, the double-peaked narrow emission lines in active galactic nuclei (AGNs) have been 
reported by Heckman et al. (1981, 1984), and they suggested biconic outflows as an origin of double 
peaked AGNs. Greene \& Ho (2005) pointed out that there are about 1\% local AGNs have double-peaked
narrow emission lines. Since narrow emission lines are generally believed to be produced by clouds 
in the narrow line regions (NLRs), they suggested that $\sim$1\% AGNs have disk-like NLRs as a simplest 
explanation. The $\sim$1\% double-peaked AGN fraction have also been found by Zhou et al. (2004) 
and Wang et al. (2009).  Especially, the flux ratios of the two peaks of  \Oiii$\lambda$ 5007 are anti-correlated 
with the ratios between their shifts relative to the host galaxies statistically, this phenomenon could be understood 
as a natural result of Kepler rotation of dual AGNs, pointing to another possible explanation of double-peaked 
narrow emission lines (Wang et al. 2009).

In the last two years, more and more attention has been paid to the study of dual AGNs (e.g. Comerford \etal 2009a, 2009b; Dotti et al. 2009; Boroson \& Lauer 2009; Xu \& Komossa 2009; Liu et al. 2010) since on the one hand, they are unavoidable results of hierarchical cosmology model
which has gained great success; on the other hand, with the launch of X-ray and infrared space astronomical facilities (Chandra,  Spitzer, and Herschel Space telescopes), it is now possible for us to resolve the dual
AGNs with kpc-scale separation. Observational evidence for dual AGNs includes: spatially resolved systems 
in which both supermassive black holes (SMBHs) can be identified directly and spatially unresolved systems in which 
the dual AGN model can explain various phenomena (see Komossa 2006 for a detail review). Most recently, Colpi \& Dotti 2009 further summarize the observations and numerical simulations of dual and binary black holes. So far, 
a few unambiguous cases have been found, such as NGC 6240 (Komossa \etal 2003), J0402+379 (Rodriguez et al. 2006, 
2009; Morganti et al. 2009), EGSD2 J142033+525917 (Gerke \etal 2007), EGSD2 J141550+520929 
(Comerford \etal 2009a), COSMOS J100043+020637 (Comerford \etal 2009b), and other four dual AGNs (see Liu et al. 2010). 
There are other interesting sources, which need more certification by future observations. Zhou et al. (2004) connected both SDSS and VLBA 
data, suggesting SDSS J1048+0055 is a dual AGN system and double-peaked narrow emission lines could be an 
effective way of selecting dual AGN candidates. Xu \& Komossa (2009) analyze the line structures and flux ratios of 
SDSS J1316+1753 in detail, discussing all the possible origins of the double peaks. Furthermore, 
SDSS J1536+0441 is the only source in which two broad line systems have been found, and it is 
suggested to be a binary black hole system which is separated by 0.1pc with a orbital period of 100 yr (Boroson \& Lauer 2009). 
However it is also the most controversial case, which brings about a vast deal of debate 
(Chornock et al. 2009, 2010; Wrobel \& laor 2009; Decarli et al. 2009; Lauer \& Boroson 2009; Tang \& Grindlay 2009; 
Gaskell 2010; Dotti \& Ruszkowski 2010; Bondi \& P\'{e}rez-Torres 2010). 

The peculiar emission-line spectrum of SDSS J1425+3231 is noticed in the course of searching for dual AGN candidates 
in the SDSS QSO sample. SDSS J1425+3231 shows all its strong narrow emission lines with double-peaked, it is hard to make 
conclusions on the structures of weak emission lines due to the signal-to-noise ratio (S/N). We analyze the spectra of 
SDSS J1425+3231 in \S2. The black hole mass is estimated in \S3. The possible origins of the 
double-peaked line profiles of this source are discussed in \S4. The results are summarized in \S5. Throughout this paper, a cosmology with 
{\slshape H}$_0$= 70 \kms Mpc$^{-1}$, $\Omega_M$= 0.3, and $\Omega_\Lambda$= 0.7 is adopted. 


\section{Data analysis}
\label{sect:data}
SDSS J1425+3231 is a broad line QSO with SDSS pipeline redshift of $z=0.478$, all its strong narrow emission lines show 
double-peaked profiles.  In this section, we describe the procedure of spectral fitting. The steps of our analysis 
are as follows: (1) the spectrum is corrected for foreground Galactic extinction and shifted to the rest-frame by using $z=0.478$; (2) the continuum 
of the spectrum is modeled by three components (Hu et al. 2008) and subtracted, the aim is to separate the contribution of 
continuum and emission line spectrum; (3) multiple Gaussian components are used to fit the emission lines. Step (2) and (3) are 
described in more detail below.

\subsection{Continuum decomposition}
The continuum is modeled as 
\begin{eqnarray}
  \label{equ-conti}
  F_\lambda & = & F_\lambda^{\rm PL}(F_{5100},\alpha)
  +F_\lambda^{\rm BaC}(F_{\rm BE},\tau_{\rm BE}) +F_\lambda^{\rm Fe}(F_{\rm Fe}, {\rm FWHM_{\rm Fe}}, V_{\rm Fe}).
\end{eqnarray}
where $F_\lambda^{\rm PL}=F_{5100} (\frac{\lambda}{5100}) ^{\alpha}$ is a featureless power law, $F_{5100}$ is the 
flux at 5100 \AA\ and $\alpha$ is the spectral index. The second and third terms represent the Balmer continuum and 
Fe emission, respectively.

For wavelength shortward of the Balmer edge $\lambda < \lambda_{\rm BE}=3646$\AA, the Balmer continuum can be expressed 
as $F_\lambda^{\rm BaC}=F_{\rm BE}B_\lambda(T_e)(1-e^{-\tau_\lambda})$ (Grandi 1982; Dietrich et al. 2002). $F_{\rm BE}$ 
is a normalization coefficient for the flux at $\lambda_{\rm BE}$, $B_\lambda(T_e)$ is the Planck function at an electron 
temperature $T_e$, $\tau_\lambda=\tau_{\rm BE}( \frac{\lambda}{\lambda_{\rm BE}})$ is the optical depth at wavelength 
$\lambda$, $\tau_{\rm BE}$ is the optical depth at the Balmer edge. $T_e$ is assumed to be $T_e=15,000$K. The two free
parameters in the Balmer continuum are $F_{\rm BE}$ and $\tau_{\rm BE}$. At $\lambda>\lambda_{\rm BE}$, blended higher-order 
Balmer lines give a smooth rise from $\sim$4000 \AA\ to the Balmer edge (Wills et al. 1985) in the spectrum. However, 
our fitting windows do not include this region, actually our results are not influenced by the higher order Balmer lines.

The optical and ultraviolet Fe II template ($F_\lambda^{\rm I Zw 1}$) from NLS1 I ZW 1 is used to subtract the Fe II emission 
from the spectra (Boroson \& Green 1992; Vestergaard \& Wilkes 2001). The I ZW 1 template is broadened by convolving with a 
Gaussian function $G$:
\begin{equation}
  \label{equ-fe}
  F_\lambda^{\rm Fe}=F_\lambda^{\rm I Zw 1}\ast G(F_{\rm conv}, {\rm
  FWHM_{conv}}, V_{\rm conv}),
\end{equation}
where $F_{\rm conv}$, ${\rm FWHM_{conv}}$, and $V_{\rm conv}$ are the flux, width and peak velocity shift of the Gaussian 
function. The parameters of Fe in equation (1) can be expressed as follows: the flux of the Fe emission, $F_{\rm Fe}$, it is 
the multiplication of $F_{\rm conv}$ and the flux of the template;  the shift of the Fe spectrum, $V_{\rm Fe} = V_{\rm conv}$, 
and the FWHM of the Fe lines ${\rm FWHM_{\rm Fe}}=\sqrt{{\rm FWHM_{I Zw 1}^2+FWHM_{conv}^2}}$.

   \begin{figure}[h!!!]
   \centering
   \includegraphics[width=15.0cm, angle=-90]{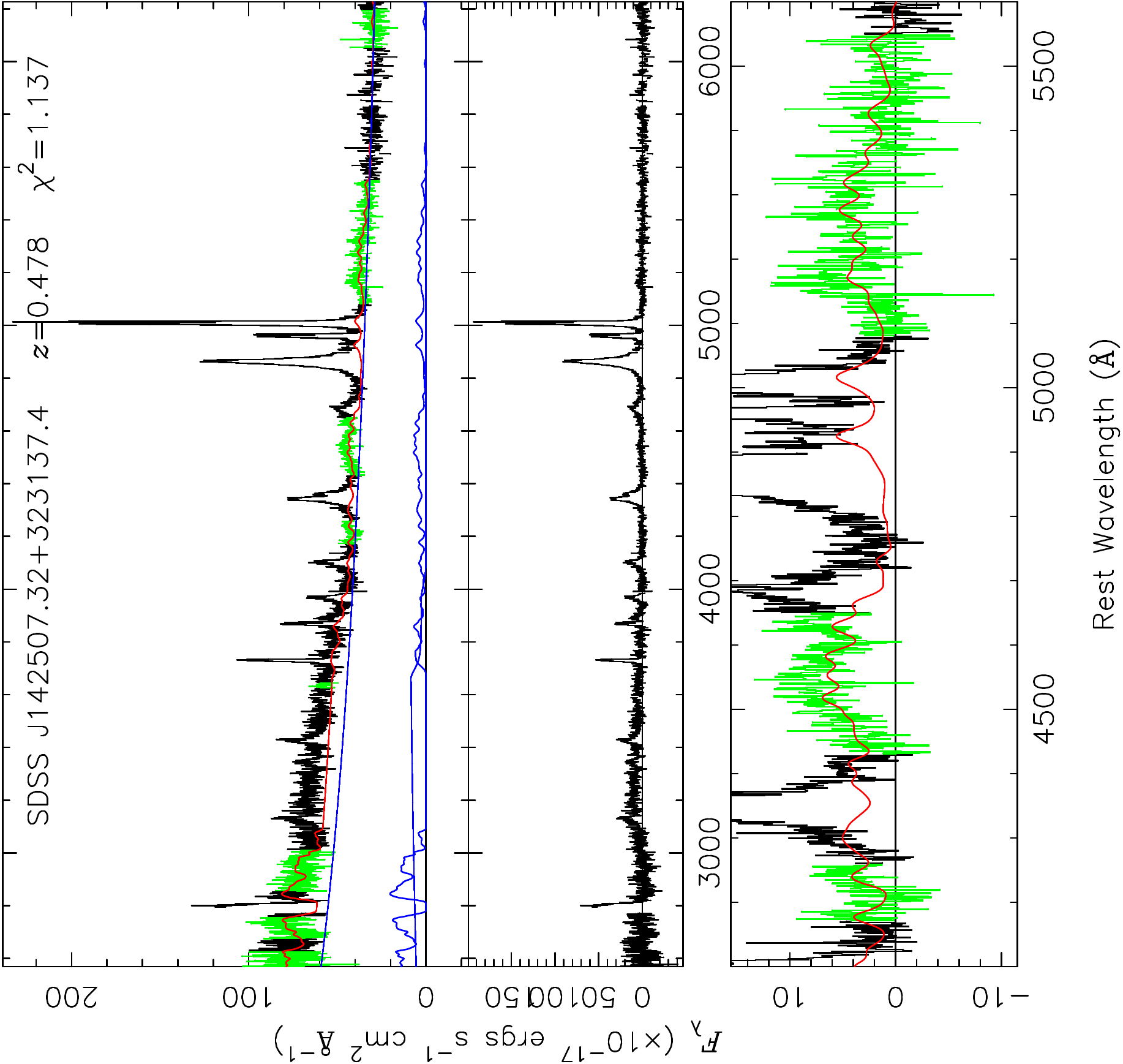}
   \begin{minipage}[]{15cm}
   \caption{ SDSS spectrum of SDSS J1425+3231 plotted as observed flux versus rest wavelength. The top panel shows the 
   Galactic extinction and redshift corrected spectrum. The spectrum in the fitting window is plotted in green. The three components 
   of the continuum are shown in blue while The best fit model is shown in red. The middle panel shows the residual spectrum, 
   namely the pure emission-line spectrum. The bottom panel shows the spectrum after subtracting the power law and the Balmer 
   continuum in the wavelength range 4100--5600 \AA. The Fe model is shown in red. The flux density $F_\lambda$ is given in units 
   of 10$^{-17}$ erg s$^{-1}$cm$^{-2}$\AA$^{-1}$ } \end{minipage}
   \label{Fig1}
   \end{figure}

In total, there are seven parameters in the continuum model, they are fitted by minimizing $\chi^2$. The fitting windows 
include: 2470--2625, 2675--2755, 2855--3010, 3625--3645, 4170--4260, 4430--4650, 5080--5550, and 6050--6200 \AA. 
These windows are free of strong contaminant lines. Figure~1 shows the result of the continuum decomposition. 
The Galactic extinction and redshift corrected spectrum is shown in the top panel. The spectrum in the fitting window is plotted in 
green. The three components of the continuum are shown in blue. The best fit model is shown in red. The middle panel shows the 
residual spectrum, namely the pure emission-line spectrum. We will analyze it in the next step. In the bottom panel, we subtract the 
power law and the Balmer continuum, zoom in Fe-only spectrum in the wavelength range 4100--5600 \AA. The Fe model is 
shown in red.  

   \begin{figure}[h!!!]
   \centering
   \includegraphics[width=15.0cm, angle=0]{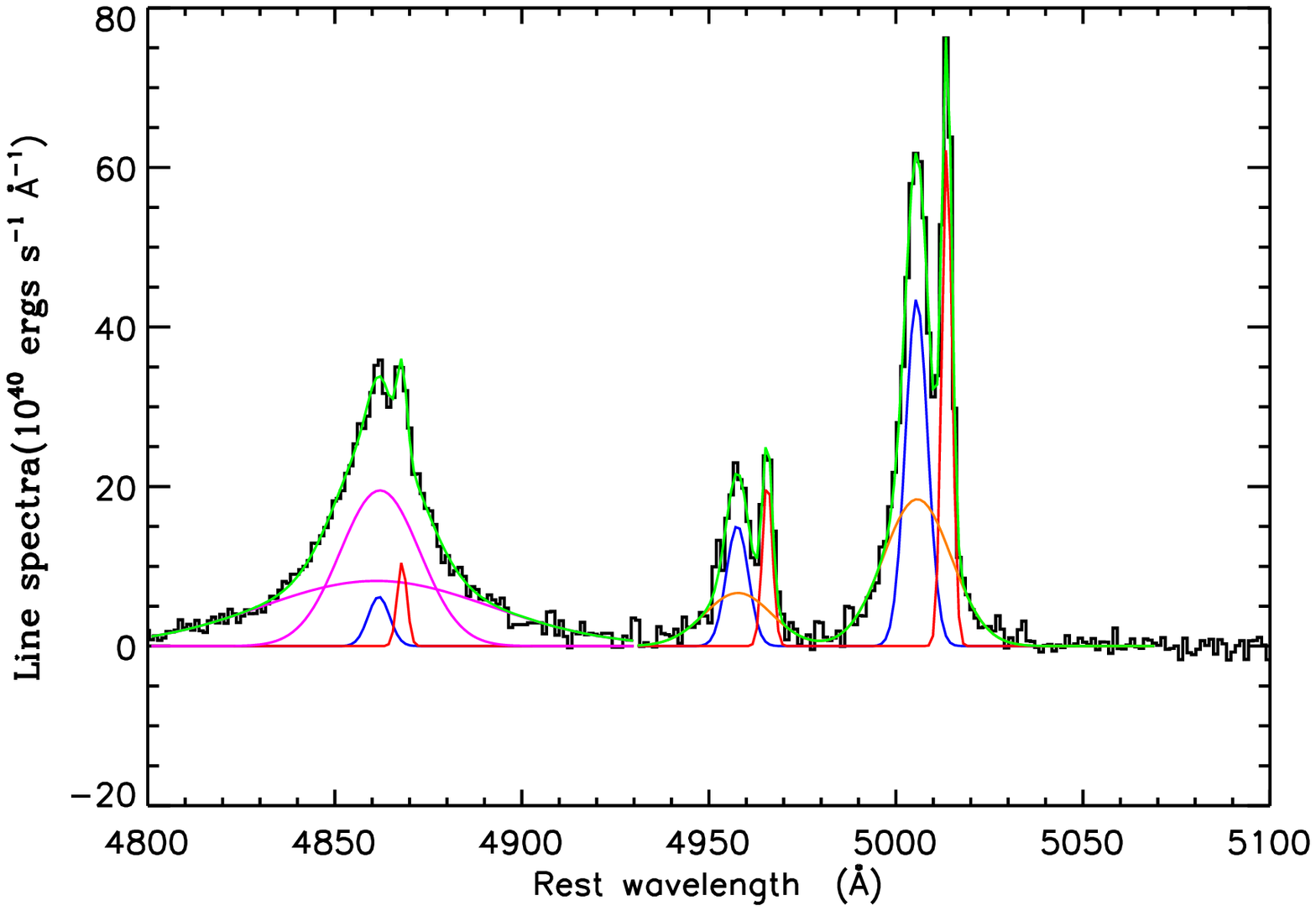}
   \begin{minipage}[]{15cm}
   \caption{ Emission-line fitting of SDSS J1425+3231.  For \Oiii~$\lambda$4959, 5007 emission lines, 
  three Gaussian components are used. The blue and red Gauss represent the two narrow line peaks, 
  the orange is the underlying  wing. The broad H$\beta$ are fitted 
  with four Gaussian components, two for the double-peaked narrow line (red and blue) and another two 
  for the broad line (see the two pink components). The best fit model is shown in green.} \end{minipage}
   \label{Fig2}
   \end{figure}

\subsection{Emission line fitting}
We measured the H$\beta$ and \Oiii~$\lambda$4959, 5007 emission lines from the emission line spectrum. 
Since the narrow lines of this object are double-peaked, two sets of three Gaussian components are used to 
model \Oiii~$\lambda$4959 and 5007. One for the blue component, one for the red component, and one extra Gaussian 
component for the underlying broad wing. Fit parameters of all the three Gaussians are the central wavelength, $\sigma$ 
and flux. Each \Oiii~$\lambda$4959 component is forced to have the same centroid and width as the corresponding
\Oiii~$\lambda$5007. In Figure~2, the blue and red components are shown in blue and red respectively. The orange is the 
underlying wing. The best fit model is shown in green. The three-Gaussian parameterization describes the observed 
\Oiii~$\lambda$4959, 5007 lines very well. The blue and red components are separated by $\sim$500\kms in velocity space, 
and the $\sigma$ of blue component is 417\kms, about 2.2 times broader than the red component. The broad wing (orange) peaks at 
roughly the same position as that of the blue component. The $\sigma$ of the wing is 628\kms. We use 
four-Gaussian components to model the asymmetric profile of the H$\beta$ emission, two for the blue and red narrow peaks, 
they are forced to have the same width as the corresponding \Oiii~$\lambda$5007, see the blue and red Gauss 
in Figure~2. Different from the fit of \Oiii, we do not fit a wing to the narrow H$\beta$ component since on the one 
hand, comparing with the broad H$\beta$, the strength of the wing can be neglected; on the other hand, the 
resolution of the SDSS spectrum is not high enough to separate such a weak component from the broad H$\beta$. 
The broad H$\beta$ are fitted with the two other Gaussian components, see the pink lines. 
The fitting parameters of each Gaussian component is shown in table 1.

\begin{table}
\bc
\begin{minipage}[]{100mm}
\caption[]{ Emission line properties of SDSS J1425+3231}\end{minipage}
\small
\begin{tabular}{lcccccccccc}
  \hline\noalign{\smallskip}
  \hline\noalign{\smallskip}
Property & H$\beta$ & \Oiii  &  \Oiii \\
   & 4861.33 & 4958.91 & 5006.84  \\
  \hline\noalign{\smallskip}
Blue system &  &  &  \\
\ \ \ \ Line center (\AA) & 4861.77$\pm$0.51 & 4957.59$\pm$0.08 & 5005.51$\pm$0.08 \\
\ \ \ \ FWHM (\kms) & 417.17$\pm$17.77 & 417.17$\pm$17.77 & 417.17$\pm$17.77 \\
\ \ \ \ Luminosity$^a$ & 44.51$\pm$9.88 & 110.98$\pm$11.40 &  322.76$\pm$24.22 \\
  \hline\noalign{\smallskip}
Red system &  &  &  \\
\ \ \ \ Line center (\AA) & 4867.97$\pm$0.19 & 4965.64$\pm$0.04 & 5013.64$\pm$0.04 \\
\ \ \ \ FWHM (\kms) & 188.21$\pm$5.60 & 188.21$\pm$5.60 & 188.21$\pm$5.60 \\
\ \ \ \ Luminosity$^a$ & 34.33$\pm$4.48 & 70.31$\pm$4.44 & 212.61$\pm$7.59 \\
  \hline\noalign{\smallskip}
Wing &  &  &  \\
\ \ \ \ Line center (\AA) & - & 4957.78$\pm$0.31 & 5005.70$\pm$0.31 \\
\ \ \ \ FWHM (\kms) & - & 1281.42$\pm$58.31 & 1281.42$\pm$58.31 \\
\ \ \ \ Luminosity$^a$ & - & 150.33$\pm$16.83 & 419.44$\pm$27.07 \\  
  \hline\noalign{\smallskip} 
Broad system &  &  &  \\
\ \ \ \ Line center (\AA) & 4862.09$\pm$0.28 & - & - \\
\ \ \ \ FWHM (\kms) & 1545.02$\pm$82.42 & - & - \\
\ \ \ \ Luminosity$^a$ & 520.66$\pm$38.74 & - & - \\
Very broad system$^b$ &  &  &  \\
\ \ \ \ Line center (\AA) & 4861.08$\pm$0.96 & - & - \\
\ \ \ \ FWHM (\kms) & 4463.45$\pm$268.77 & - & - \\
\ \ \ \ Luminosity$^a$ & 629.68$\pm$38.88 & - & - \\
  \noalign{\smallskip}\hline
\end{tabular}
\ec
\tablecomments{0.86\textwidth}{\\ $^a$ In units of $10^{40}$ erg s$^{-1}$.\\
$^b$ Note only the H$\beta$ need the very broad component in appearance.}
\end{table}


\section{Mass of the Central Black Hole}
\label{sect:mass}
In this section, we use three different methods to estimate the mass of the central black hole.

The first method we used is based on the virial theorem and $R_{\rm BLR}-L_{5100}$ relation which is calibrated with the reverberation mapping data. 
The BLR radius $R_{\rm BLR}$ is estimated from the continuum luminosity at 
5100\AA~ ($L_{5100}=1.6\times 10^{40} {\rm erg~s^{-1}~\AA^{-1} }$) using the $R_{\rm BLR}$ $-$ $L_{5100}$ relation 
given by Bentz et al. (2006), then the black hole mass is estimated from $M_{\rm BH}= f \frac{R_{\rm BLR}\Delta V^2}{G}$, 
where $f$ is the scaling factor which is introduced to characterize the unclear kinematics and geometries 
of the BLRs. Actually, the value of $f$ changes with the shape of the line in use,
$f=3.85$ is a mean value suggested by Collin et al. (2006).
$\Delta V=1325 {\rm km~s^{-1}}$ is the second moment of the BLR H$\beta$ profile, which is reconstructed from the two pink 
components in Figure 2. This method gives a value of $8.5\times10^7M_\odot$. 

The $M_{\rm BH} - L_{5100}$ relation (see Eq. 9 of Peterson et al. 2004) predicts a mass of  $3.50\times10^{8}M_\odot$.
We also use velocity dispersion of gas in the NLR, namely $\sigma$ of  \Oiii$\lambda$ 5007 as a surrogate of galaxy stellar 
velocity dispersion $\sigma_*$ (Nelson 2000), to estimate the central black hole mass based on the famous $M_{\rm BH} - \sigma_*$ relation 
which is in the form of ${\rm{log}}({M_{\rm BH}}/{M_\odot}) = 8.13 + 4.02 {\rm{log}}({\sigma_*}/{200})$(Tremaine et al. 2002). The 
 blue component gives a mass of $8.33\times10^{7}M_\odot$ with $\sigma=177.40~$\kms ~ while the red component indicates 
 a mass of $3.42\times10^{6}M_\odot$ with $\sigma=80.18~$\kms.

\section{Discussion}
\label{sect:discussion}

\subsection{Superposition of two objects}
This possibility is very unlikely for two reasons. First, the red component of  \Oiii$\lambda$ 5007 has a luminosity of
$3.2\times10^{42}$ erg s$^{-1}$, while the $\sigma$ of this line indicates a black hole mass of $3.42\times10^6M_\odot$.
If this component comes from a background AGN, its accretion rate should be extremely super-Eddington, which is 
hard to explain (see section 4.3 for the explanation under the dual AGN scenario). Second, 
Dotti \& Ruszkowski (2010) examined the superposition model of double-peaked emission line AGNs based on galaxy 
clusters from the Millennium Run, finding that the fraction of superimposed galaxy pairs peaks at about $z=0.2$ and 
decreases rapidly since $z=0.3$. Considering the redshift of SDSS J1425+3231, $\sim 0.478$, the possibility of 
superposition should be very low.




\subsection{NLR kinematics}
Another possible explanation for the double-peaked narrow emission could be special NLR geometries such as 
biconical outflows and disk-like NLRs. In such hypothesis, there is only one AGN to illuminate the NLR gas which is 
moving toward and away from us, forming the blue and red components in the observed narrow lines.

Certain nearby Seyfert and star forming galaxies are known to have biconic outflow induced double-peaked 
emission lines. The examples are found not only from the spatial resolved spectra which takes along the minor axis 
of a galaxy (e.g. Cecil 1988; Cecil et al. 1990; Veilleux et al. 1994, 2001; Colbert et al. 1996) but also from the spectrum of the whole galaxy (e.g. Duric \& Seaquist 1988; Axon et al. 1998). In the 
scenario of biconic outflows, we would expect that the blue and 
red components have the same velocity dispersion since they are illuminated by the same AGN.  However, 
in SDSS J1425+3231, the blue component is three times broader than the red component, which is conflicted with the 
outflow model.  On the other hand, the outflow studies find that the NLRs are stratified strongly in ionization and 
velocity so that high-ionization lines, such as \Oiii\ are originated near the AGN with higher velocity and low-ionization 
lines, such as \Hb\ and \Oii, are originated further from the AGN with lower velocity (e.g., Komossa et al. 2008). 
In SDSS J1425+3231, we find that \Oiii\ and \Hb\ are consistency in velocity offsets within error bars,  and 
apparently no ionization stratification is observed as expected in AGN driven outflows.
In addition, the \Oiii$\lambda$5007 luminosity is $4.3 \times10^{42} \rm{erg~s}^{-1}$ for the blue system and 
$2.8 \times10^{42} \rm{erg~s}^{-1}$ for the red system. These are typical values for the emission from the whole 
NLRs of bright Seyfert galaxies and quasars, but it is hard to image that the whole NLRs are outflowing.

The observed double-peaked narrow emission lines can be accounted for in a disk-like NLR model. 
The rotating disk model predicts that the blue and red components have similar width, which is conflict with 
the current data. Furthermore, the red and blue components are expected to be (almost) equally shifted with respect to
the true cosmological galaxy redshift in this scenario. We are lack of host galaxy information since it is over-shined 
by the central AGN,  however under the disk-like NLR model, the broad line should be at the redshift of the galaxy, in
between the blue and red components, and not so close to the the blue component as we have seen from the data.

\subsection{A dual AGNs system }

The final picture we want to suggest for SDSS J1425+3231 is that this is a dual AGN system. The primary
black hole in this system has a mass of $\sim 10^8M_\odot$, the observed blue narrow component 
 in \Oiii$\lambda\lambda$4959, 5007, the broad \Hb\ component and the wing of  \Oiii$\lambda\lambda$4959, 5007 are generated in its NLR, 
 BLR and the region in between (this region is usually referred as intermediate line region in literatures) respectively. The consistency between the black hole masses estimated from 
 broad \Hb\ component and $\sigma$ of the blue component supports this idea. We have not observed the broad 
 emission lines from the BLR of the secondary black hole ($\sim 10^6M_\odot$),  the possible reasons for this 
 includes: (1) the S/N  of SDSS spectra is not high enough for us to separate the broad emission line of the 
 secondary black hole from primary one; (2) the secondary black hole is a type 2 AGN in which the BLR is 
 obscured by the dusty torus.
 
 The only issue about the dual AGN system is that, for the secondary black hole,  comparing with the 
 \Oiii$\lambda$ 5007 luminosity, a mass of $10^6M_\odot$ is a little bit low. This indicates the secondary 
 black hole is accreting in a super Eddington regime. At first sight, this is conflicted with the normal 
 accretion theory. However, we should note that in the dual AGN case, the separation between the two black
 holes is in kpc scale, the NLR gas of 
the secondary black hole can be affected by the primary black hole, namely the primary black hole can 
illuminate the NLR of the secondary black hole.

\section{Conclusion}
\label{sect:conclusion}

Type 1 QSO SDSS J1425+3231 has double-peaked narrow emission lines. In this paper, we 
analyze the SDSS spectrum of this object, discussing the origins of its double-peaked line
structure. We argue against the possibility of superposition of two objects, biconic outflow and disk like NLR, suggesting that this is a dual AGN system. 

In this system, the primary black hole has a mass of  $\sim 10^8M_\odot$, the observed 
broad lines belong to it. The secondary black hole is much smaller, its mass is in the oder of 
$10^6M_\odot$. The secondary black hole could be a type 2 AGN whose BLR is obscured, or we failed to separated the broad emission line of the secondary black hole from that of the primary black hole due to the resolution of the SDSS spectra. 

In the future observation, high spatial resolution two-dimensional optical spectrum, and imaging in the optical, radio, X-ray would help us figure out whether SDSS J1425+3231 contains dual AGNs. Moreover, the ultimate confirmation or rejection of the dual AGN interpretation, which predict the variations in the line profiles, will come from 
multi-epoch spectroscopic monitoring.

\normalem
\begin{acknowledgements}
We thank the anonymous referee for suggestions that led to improvements
in this paper. The research is supported by the National Natural Science Foundation of China
 (NSFC) under NSFC-10878010, 11003007 and 10633040, 
and  the National Basic Research Program 
 (973 program No. 2007CB815405).
 This research has made use of NASA's Astrophysics Data System Bibliographic
 Services and the NASA/IPAC Extragalactic Database (NED) which is operated
 by the Jet Propulsion Laboratory, California Institute of Technology, under
 contract with the National Aeronautics and Space Administration.
 This work is based on observations made with the {\it Spitzer Space Telescope},
 which is operated by the Jet Propulsion Laboratory, California Institute
 of Technology, under NASA contract 1407.
 
 Funding for the creation and distribution of the SDSS Archive has
been provided by the Alfred P. Sloan Foundation, the Participating
Institutions, the National Aeronautics and Space Administration, the
National Science Foundation, the US Department of Energy, the
Japanese Monbukagakusho, and the Max Planck Society. The SDSS Web
site is http://www.sdss.org.

The SDSS is managed by the Astrophysical Research Consortium (ARC)
for the Participating Institutions. The Participating Institutions
are The University of Chicago, Fermilab, the Institute for Advanced
Study, the Japan Participation Group, The Johns Hopkins University,
Los Alamos National Laboratory, the Max- Planck-Institute for
Astronomy (MPIA), the Max-Planck-Institute for Astrophysics (MPA),
New Mexico State University, University of Pittsburgh, Princeton
University, the United States Naval Observatory and the University
of Washington.
\end{acknowledgements}



\label{lastpage}

\end{document}